\documentstyle[twoside,fleqn,espcrc2,psfig]{article}

\newcommand{\puncspace}{\ifmmode\,\else{\ifcat.\C{\if.\C\else%
\if,\C\else\if?\C\else\if:\C\else\if;\C\else\if-\C\else%
\if)\C\else\if/\C\else\if]\C\else\if'\C%
\else\space\fi\fi\fi\fi\fi\fi\fi\fi\fi\fi}%
\else\if\empty\C\else\if\space\C\else\space\fi\fi\fi}\fi}%
\newcommand{\SP}{\let\\=\empty\futurelet\C\puncspace}
\newcommand{\dash}{\hbox{--}}
\newcommand{\Hz}{{\hbox{Hz}}\SP}
\newcommand{\km}{{\hbox{km}}\SP}
\def\fu#1{\leavevmode\hbox{4U~#1}\SP}
\newcommand{\nonrot}{{\rm 0}}
\def\lta{\ifmmode {\,\mathbin{\lower 3pt\hbox   
    {$\,\rlap{\raise 5pt\hbox{$\char'074$}}\mathchar"7218\,$}}}
    \else {${\mathbin{\lower 3pt\hbox
    {$\rlap{\raise 5pt\hbox{$\char'074$}}\mathchar"7218\,$}}}
    $}\fi}

\title{Constraints on the Equation of State of Neutron
Star Matter From Observations of Kilohertz QPOs}

\author{M. Coleman Miller\address{University of Chicago,
        Department of Astronomy and Astrophysics,
        5640 S. Ellis Avenue\\
        Chicago, IL  60637,
        USA}
        Frederick K. Lamb\address{University of Illinois
        at Urbana-Champaign,
        Department of Physics and Department of Astronomy\\
        1110 W. Green St.,
        Urbana, IL  61801,
        USA}
        and
        Dimitrios Psaltis\address{Harvard-Smithsonian
        Center for Astrophysics,
        60 Garden St.,
        Cambridge, MA 20218,
        USA}}

\begin{document}

\begin{abstract}
The frequencies of the highest-frequency kilohertz QPOs
recently discovered in some sixteen neutron stars in low-mass
X-ray binary systems are most likely the orbital
frequencies of gas in Keplerian orbit around these
neutron stars. If so, these QPOs provide tight upper
bounds on the masses and radii of these neutron stars and
important new constraints on the equation of state of
neutron star matter. If the
frequency of a kilohertz QPO can be established
as the orbital frequency of gas at the innermost stable
circular orbit, this would confirm one of the key predictions of
general relativity in the strong-gravity regime. If 
the spin frequency of the neutron star can also be determined, the
frequency of the QPO would fix the mass of the neutron
star for each assumed equation of state. Here we show how
to derive mass and radius bounds, using the kilohertz
QPOs, for nonrotating and slowly rotating stars, and
discuss how these bounds are affected by rapid stellar
rotation and radial radiation forces. We also describe 
observational results that
would be strong evidence for the presence of an innermost 
stable circular orbit. No such strong evidence is present
in current data, but future prospects are excellent.
\end{abstract}

\maketitle

\section{INTRODUCTION}

Determination of the equation of state of neutron
stars has been an important goal of nuclear physics for
more than two decades. Progress toward this goal can be
made by establishing astrophysical constraints as well
as by improving our understanding of nuclear forces.

Many ways of deriving constraints from astrophysics have
been explored. One of the best known is pulse timing of
pulsars in binary systems. Although binary pulsar timing
has made possible stringent tests of general relativity
(see, e.g., [16]), the $\approx 1.4\,M_\odot$
masses derived from timing (see [17])
are allowed by all equations of state based on realistic
nuclear physics, and hence these observations have not
eliminated any of the equations of state currently being
considered. The highest known neutron star spin
frequency, the 643~Hz frequency of PSR~1937$+$21
[1], is
also allowed by all equations of state currently under
consideration. Radius estimates based on the energy
spectra of type~I X-ray bursts and on observations of
thermal emission from the surfaces of neutron stars are
more restrictive in principle but currently have large
systematic uncertainties (see [6,11]).

The discovery of high-frequency brightness oscillations
from some sixteen neutron stars in low-mass X-ray binaries
with the {\it Rossi} X-ray Timing Explorer (RXTE) holds
great promise for providing important new constraints.
Oscillations are observed both in the persistent X-ray
emission and during type~I X-ray bursts. The kilohertz
quasi-periodic oscillations (QPOs) observed in the
persistent emission have high amplitudes and relatively
high coherences (see, e.g., [18]). A pair of
kilohertz QPOs is commonly observed in a given source.
Although the frequencies of these QPOs vary by up to a
factor of $\sim$2, the frequency separation $\Delta\nu$
between a pair of kilohertz QPOs appears to be constant
in almost all cases. In both the sonic-point [10]
and magnetospheric [15]
beat-frequency interpretations, the higher
frequency in a pair is the orbital frequency at the
inner edge of the Keplerian flow, whereas the lower
frequency is the beat of the stellar spin frequency with
this frequency. Such high orbital frequencies yield
interesting bounds on the masses and radii of these
neutron stars and interesting constraints on the
equation of state of neutron star matter.
In \S~2 we describe how mass and radius bounds can be
derived from the properties of
the kilohertz QPOs, and discuss how these bounds are affected by
rotation and by radial radiation forces. 

The bounds on the mass and radius and on the equation of state
would become particularly restrictive if the frequency 
of a kilohertz QPO can be
securely established as the orbital frequency of gas
at the innermost stable circular orbit. Indeed, if the QPO
frequencies currently observed in some sources are
the orbital frequency of gas at the innermost
stable orbit, several currently viable equations of
state are ruled out. Moreover, detection of the effects
of the innermost stable orbit would, by itself, be 
a confirmation of one
of the key predictions of general relativity in the
strong-gravity regime. Several authors have recently suggested
that these effects have already been observed and have
argued that the neutron stars in
some kilohertz QPO sources therefore have masses close to
$2.0\,M_\odot$. In \S~3 we discuss these suggestions, and
describe observational results that
would be strong evidence for the presence of an
innermost stable orbit.
We argue that the evidence cited to support detection of the
innermost stable circular orbit was
not compelling and that more recent observational results do not support
these claims.

\section{CALCULATIONS}

Suppose that, as in the sonic-point model, the
frequency $\nu_{\rm QPO2}$ of the higher-frequency
QPO in a kilohertz pair is the orbital frequency of
gas in a nearly circular Keplerian orbit around the
neutron star and that the highest observed value of
$\nu_{\rm QPO2}$ from a given star is $\nu_{\rm QPO2}^\ast$.

\subsection{Nonrotating stars}

Assume first that the star is not rotating and is
spherically symmetric. Then the exterior spacetime
is the Schwarzschild spacetime and
the orbital frequency (measured at infinity) of gas
in a circular orbit at Boyer-Lindquist radius $r$
around a star of mass $M$ is
\begin{equation}
 \nu^\nonrot_{\rm K}(M,r)=(1/2\pi)(GM/r^3)^{1/2}\;.
\end{equation}
Here and below the superscript 0 indicates that
the relation is that for a nonrotating star.
 Equation~(1) may be solved for
the mass of the star as a function of the radius of
the orbit with frequency $\nu_{\rm QPO2}^\ast$,
with the result
\begin{equation}
 M^\nonrot(R_{\rm orb},\nu_{\rm QPO2}^\ast) =
 (4\pi^2/G) R_{\rm orb}^3 (\nu_{\rm QPO2}^\ast)^2\;.
\end{equation}
The mass of the star
is related to the radius of the innermost
stable orbit by the expression
\begin{equation}
 M^\nonrot(R_{\rm ms}) = (c^2/6G)R_{\rm ms}\;.
\end{equation}
The
radius of the star must be smaller than the
radius $R_{\rm orb}$ of the gas with orbital
frequency $\nu_{\rm QPO2}^\ast$, so the
representative point of the star in the
$R$,$M$ plane must lie to the left of the curve
$M^\nonrot(R_{\rm orb},\nu_{\rm QPO2}^\ast)$. In
addition, in order to produce a wave train with
tens of oscillations, the gas producing the QPO
must be outside the radius $R_{\rm ms}$ of the
innermost stable circular orbit, so the
representative point must also lie below the
intersection of $M^\nonrot(R_{\rm orb},\nu_{\rm
QPO2}^\ast)$ with the curve $M^\nonrot(R_{\rm
ms})$. Figure~1a shows the allowed region of the
$R$,$M$ plane for $\nu_{\rm QPO2}^\ast =
1220~\Hz$. The maximum allowed mass and radius
are [10]
 \begin{equation}
  M^\nonrot_{\rm max} =
  2.2\,(1000~\Hz/\nu_{\rm QPO2}^\ast)\; M_\odot
 \end{equation}
 \begin{equation}
  R^\nonrot_{\rm max} =
  19.5\,(1000~\Hz/\nu_{\rm QPO2}^\ast)\;{\rm km.}
 \end{equation}
For example, the 1220~Hz QPO observed in the atoll
source 4U~1636$-$536 would constrain the mass of this
neutron star to be less than $1.8\,M_\odot$ and
the radius to be less than 16.0 km, if it were not
rotating. Figure~1b compares the mass-radius relations
for nonrotating stars
given by five equations of state with the regions
of the radius-mass plane allowed for three values of
$\nu_{\rm QPO2}^\ast$.

\begin{figure*}
\centerline{\psfig{file=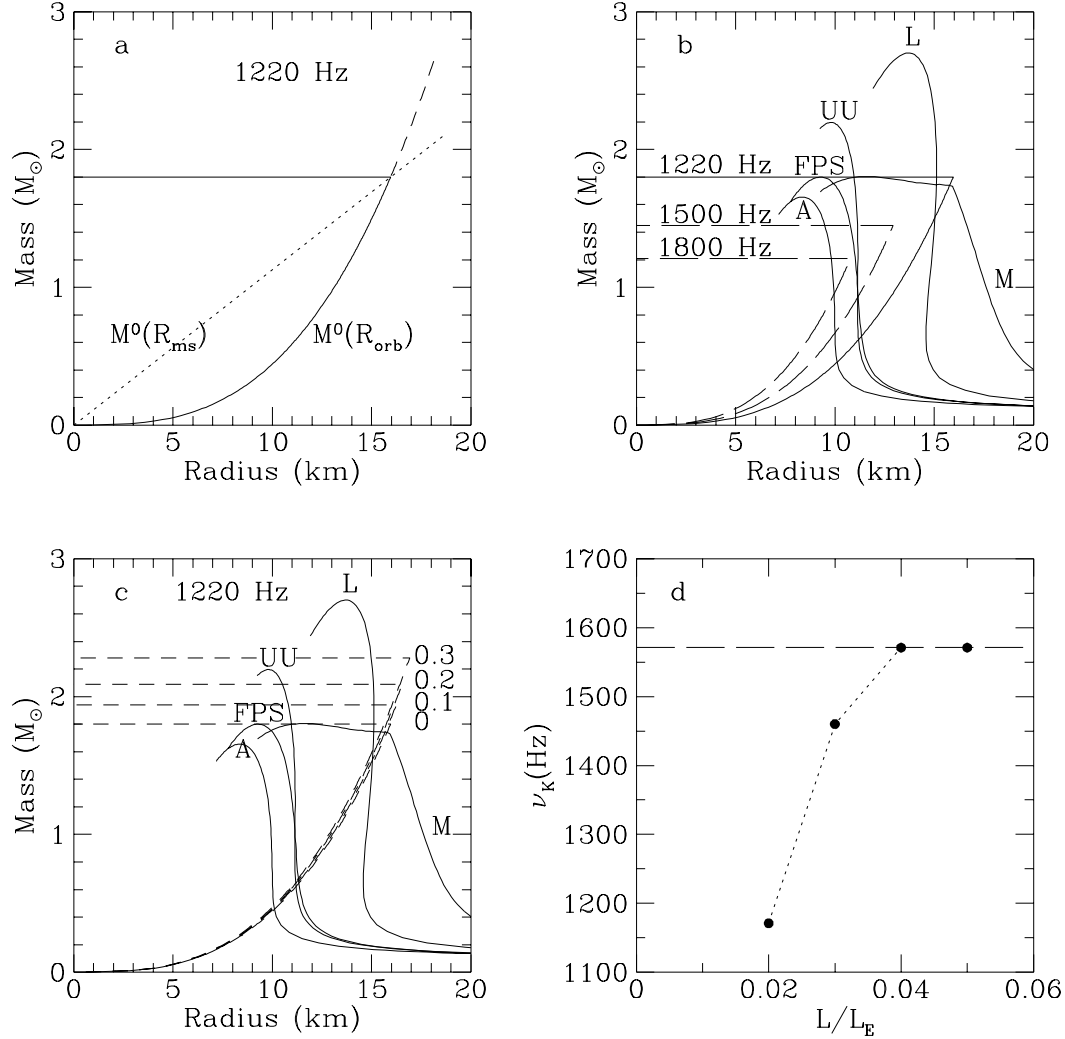,height=6truein,width=6truein}}
 \caption{
 (a)~Radius-mass plane, showing the region allowed
for a nonrotating neutron star with $\nu_{\rm
QPO2}^\ast = 1220$~Hz.
  (b)~Comparison of the mass-radius relations
for nonrotating neutron stars given by five
representative equations of state with the
regions of the mass-radius plane allowed for
nonrotating stars and three different Keplerian
orbital frequencies. The light solid curves show
the mass-radius relations given by equations of
state A [12], FPS [7],
UU [20], L [14], and M [13].
  (c)~Regions allowed for rotating neutron stars
with various values of $j$ and $\nu_{\rm
QPO2}^\ast = 1220$~Hz, when first-order
effects of the stellar spin are included.
  (d)~Illustrative Keplerian QPO frequency given
by fully general relativistic calculations of the
gas dynamics and radiation transport in the
sonic-point model [10].
 }
\end{figure*}

\subsection{Slowly rotating stars}

Rotation affects the structure of the star and the
spacetime, altering the mass-radius relation, the
frequency of an orbit of given radius, and the
radius $R_{\rm ms}$ of the innermost stable orbit.

The parameter that characterizes the importance of
rotational effects is the dimensionless quantity
\hbox{$j \equiv cJ/GM^2$}, where $J$ and $M$ are the
angular momentum and gravitational mass of the star.
The value of $j$ that corresponds to a given
observed spin frequency depends on the neutron star
mass and equation of state, and is typically higher
for lower masses and stiffer equations of state.
For the spin frequencies $\sim 300~\Hz$ inferred in
the kilohertz QPO sources, $j \sim 0.1 \dash 0.3$.
For such small values of $j$, a first-order
treatment is adequate. To this order, analytical
expressions are available for the relevant
quantities and one can prove the existence of upper
bounds on the mass and radius [10].
However, the bounds must be computed numerically.

To first order in $j$, the orbital frequency
(measured at infinity) of gas in a prograde
Keplerian orbit at a given Boyer-Lindquist radius
$r$ is
\begin{equation}
   \nu_K(r,M,j) \approx
    [1-j(GM/rc^2)^{3/2}]\,\nu^\nonrot_K(r,M)
\end{equation}
 and the radius of the innermost stable orbit is
\begin{equation}
  R_{\rm ms}(M,j) \approx
   [1-j(2/3)^{3/2}]\,R^\nonrot_{\rm ms}(M)\;,
\end{equation}
 where $\nu^\nonrot_K$ and $R^\nonrot_{\rm ms}$ are
the Keplerian frequency and radius of the innermost
stable orbit for a nonrotating star. Hence, to first
order in $j$, the frequency of the prograde orbit at
$R_{\rm ms}$ around a star of given mass $M$ and
dimensionless angular momentum $j$ is (see [4])
\begin{equation}
 \nu_{\rm K, ms} \approx
   2210\,(1+0.75j)(M_\odot/M)\,\Hz\;.
\end{equation}
As a result, for slowly rotating stars the upper 
bounds on the mass become
\begin{equation}
   M_{\rm max} \approx
   [1+0.75j(\nu_{\rm spin})]M^\nonrot_{\rm max}
\label{Mmax}
\end{equation}
 and
\begin{equation}
   R_{\rm max} \approx
   [1+0.20j(\nu_{\rm spin})]R^\nonrot_{\rm max}\;,
\label{Rmax}
\end{equation}
 where $j(\nu_{\rm spin})$ is the value of $j$ for
the observed stellar spin rate at the maximum
allowed mass for the equation of state being
considered and $M^\nonrot_{\rm max}$ and
$R^\nonrot_{\rm max}$ are the maximum allowed mass
and radius for a nonrotating star (see above).
Expressions~(\ref{Mmax}) and~(\ref{Rmax}) show
that the bounds are always greater for a slowly
rotating star than for a nonrotating star,
regardless of the equation of state assumed.

Figure~1c illustrates the effects of stellar
rotation on the region of the radius-mass plane
allowed for spin rates $\sim
300~\Hz$, like those inferred for the kilohertz QPO
sources, and \hbox{$\nu_{\rm QPO2}^\ast=1220~\Hz$},
the frequency of the highest-frequency QPO so far
observed in \fu{1636$-$536}, which is also the
highest-frequency QPO so far observed in any
source. Our calculations show that {\it the mass of
the neutron star in \fu{1636$-$536} must be less
than $\sim 2.2\,M_\odot$ and its radius must be less
than $\sim 17~\km$}. As just explained, the precise
upper bounds depend on the equation of state assumed.
For further details, see [10].

\subsection{Rapidly rotating stars}

If the stellar spin is $\sim 500~\Hz$ or higher,
spin affects the structure of the star as well
as the exterior spacetime. The exterior spacetime
of such a rapidly rotating star differs
substantially from the Kerr spacetime and must be
computed numerically for each assumed equation of
state. Derivation of bounds on the mass and radius
of a given star for an assumed equation of state
requires construction of a sequence of stellar
models and spacetimes for different masses using
the assumed equation of state, with $\nu_{\rm
spin}$ as measured at infinity held fixed. The
maximum and minimum possible masses and radii
allowed by the observed QPO frequency can then be
determined.

Such computations have been carried out by
Miller, Lamb, \& Cook [9]. They find that
if the neutron star is spinning rapidly, the
constraints on the equation of state are tightened
dramatically. For instance, if the spin frequency
of the neutron star in \fu{1636$-$536} is $\sim
580~\Hz$, the frequency of the single brightness
oscillation observed
during X-ray bursts from this source, then the
tensor-interaction equation of state of
Pandharipande and Smith [13] is ruled out by the
kilohertz QPO frequencies already observed from this source. 
They also find that observation of a 1500~Hz orbital frequency would
constrain the mass and radius of the neutron star
to be less than $\sim 1.7\, M_\odot$ and $\sim 13$~km,
ruling out several equations of state that
are currently astrophysically viable, regardless of
the star's spin rate.  

\subsection{Effects of the radial radiation force}

The luminosities of the Z sources are typically
$\sim 0.5 \dash 1\,L_E$, where $L_E$ is the
Eddington luminosity. Hence, in the Z
sources the outward acceleration caused by the
radial component of the radiation force can be a
substantial fraction of the inward acceleration
caused by gravity. 
The radially outward component of the radiation
force reduces the orbital frequency at a given
radius. For example, if the star is spherical and
nonrotating  and emits radiation uniformly and
isotropically from its entire surface, the orbital
frequency (measured at infinity) of a test particle
at Boyer-Lindquist radius $r$ is given by
 \begin{equation}
  {\nu_{\rm K}(L)\over{\nu_{\rm K}(0)}}=
  \left[1-{(1-3GM/rc^2)^{1/2}\over{(1-2GM/rc^2)}}
  {L\over{L_E}}\right]^{1/2}\;,
  \label{eq:RadForceFreq}
 \end{equation}
 where $L$ is the luminosity of the star measured at
infinity and $\nu_{\rm K}(0)$ is the Keplerian
frequency in the absence of radiation forces.
Thus, the Boyer-Lindquist radius of a circular
orbit with a given frequency is smaller in the
presence of the radial radiation force and the
constraints on the mass and radius of the star are
therefore strengthened.

For the atoll sources, which have luminosities
\hbox{$\lta 0.1\,L_E$}, the change in the
Keplerian frequency is at most $\sim 5$\%. For the Z
sources, on the other hand, which have luminosities
\hbox{$\approx L_E$}, the change may be much
larger, although the change in the sonic-point
Keplerian frequency may be smaller than would be
suggested by a naive application of
equation~(\ref{eq:RadForceFreq}), if a substantial
fraction of the radiation produced near the star is
scattered out of the disk plane before it reaches
the sonic point.

\section{INNERMOST STABLE CIRCULAR ORBIT}

Establishing that an observed QPO frequency is
the orbital frequency of the innermost stable
circular orbit in an X-ray source would be an important step
forward in our understanding of strong-field
gravity and the properties of dense matter,
because it would (1)~confirm one of the key
predictions of general relativity in the
strong-field regime and (2)~fix the mass of the
neutron star in that source, for each assumed
equation of state.

Given the fundamental significance of the
innermost stable orbit, it is very important to
establish what would constitute strong, rather
than merely suggestive, evidence that the
innermost stable orbit has been detected.
Clear signatures of the innermost stable orbit
include the following (see [10] for
more details):

(1)~Probably the most convincing signature
would be a fairly coherent, kilohertz QPO with a
frequency that reproducibly increases steeply
with increasing accretion rate but then becomes
constant and remains nearly constant as the
accretion rate increases further.
The constant frequency should always be the same
in a given source. As shown in
Figure~1d, this behavior emerges naturally from
general relativistic calculations of the
gas dynamics and radiation transport in the 
sonic-point model [5,10].

(2)~A possible signature of the innermost
stable orbit would be a simultaneous sharp drop
in the amplitudes of both QPOs in a kilohertz
QPO pair, or a sharp drop in the amplitude of the
lower-frequency QPO, at a frequency that is
always the same in a given source.

(3)~A less likely but possible signature of the
innermost stable orbit would be a steep and
reproducible drop in the coherence of both QPOs
in a pair (or in the coherence of the
higher-frequency QPO, if the lower-frequency QPO
is not visible) at a certain critical frequency
(the frequency of the innermost stable orbit),
as the frequencies of both QPOs increase
steadily with increasing accretion rate. The
critical frequency should always be the same in
a given source.

Several authors have recently suggested that
innermost stable orbits have already been
observed. Zhang et al.\ [22] suggested that the
similarity of the highest QPO frequencies seen so
far indicates that innermost stable circular orbits are
being detected. Kaaret, Ford,
\& Chen [3]3 suggested that the roughly
constant frequencies of the 800--900~Hz
QPOs seen in \fu{1608$-$52} [2] and
\fu{1636$-$536} [21] were generated by the beat
of the spin frequency against the frequency of
the innermost stable circular orbits in these sources.
This would
imply that the neutron stars in all the
kilohertz QPO sources have masses close to
$2.0\,M_\odot$. However, {\em no strong
signatures of the innermost stable circular
orbit have so far been seen in any of these
sources}.

Indeed, more recent observations of
both \fu{1608$-$52} [8] and
\fu{1636$-$536} [19] are inconsistent with
the suggestion that the QPO frequencies seen
initially are related to the frequencies of
the innermost orbits in these sources.
The 1171~Hz QPO frequency assumed by Kaaret
et al.\ [3] to be
the frequency of the innermost stable orbit in
\fu{1636$-$536} and used by them to determine
the mass of the star was later seen to increase
to 1193~Hz and then to 1220~Hz ([19]; 
W.~Zhang, personal
communication). There is as yet no evidence for
a maximum QPO frequency in \fu{1636$-$536} and
hence there is no basis for the suggestion that
an innermost stable orbit has been seen in this
source.

A recent analysis of \fu{1608$-$52} data by
M\'endez et al.\ [8] shows that this source
has twin kilohertz QPOs that vary with countrate
just like the other atoll sources. There is as
yet no evidence for a maximum QPO frequency in
\fu{1608$-$52} and hence there is no basis for
the suggestion that an innermost stable orbit
has been seen in this source, either.

Even though no strong evidence
for the effects of the innermost stable circular
orbit has yet been seen in any kilohertz QPO data, 
there is reason for optimism, because the highest QPO frequencies
observed so far are only $100\dash 200$~Hz below
the expected frequencies of the
innermost stable orbit around these neutron stars.
Given the rapid pace of discoveries in this field,
the prospects for obtaining clear 
evidence of an innermost stable circular orbit in 
the future appear excellent. 
\bigskip

This work was supported in part by NASA grant NAG~5-2868 
at the University of Chicago and by NSF grants
AST~93-15133 and AST~96-18524 and NASA grant NAG~5-2925
at the University of Illinois.


\begin{thebibliography}{9}

\bibitem{B82} Backer, D.\,C., Kulkarni, S.\,R.,
Heiles, C., Davis, M.\,M., \& Goss, W.\,M. 1982,
Nature, 300, 615

\bibitem{B96}
Berger, M., et al.\ 1996, ApJ, 469, L13

\bibitem{KFC97}
Kaaret, P., Ford, E.~C., \& Chen, K. 1997, ApJ,
480, L27

\bibitem{KMW90}
Klu\'zniak, W., Michelson, P., \& Wagoner, R.~,V.
1990, ApJ, 358, 538

\bibitem{LMP97} Lamb, F.\,K., Miller, M.\,C., \& Psaltis,
D. 1997, these proceedings

\bibitem{LVT93} Lewin, W.\,H.\,G., van Paradijs, J.,
\& Taam, R.\,E. 1993, Sp. Sci. Rev., 62, 223

\bibitem{LRP93}
Lorenz, C.~P., Ravenhall, D.~G., \& Pethick, C.~J.
1993, Phys. Rev. Lett. 70, 379

\bibitem{M97}M\'endez, M., et al. 1997, ApJ, submitted
(preprint astro-ph/9712085)

\bibitem{MLC97}
Miller, M.~C., Lamb, F.~K., \& Cook, G. 1997,
in preparation

\bibitem{MLP97}
Miller, M.~C., Lamb, F.~K., \& Psaltis, D. 1997,
ApJ, in press (astro-ph/9609157)

\bibitem{Og93} \"Ogelman, H. 1995, in Proc. of NATO
ASI on The Lives of
Neutron Stars, ed. A.\,Alpar et al.\ (Dordrecht: Kluwer), 101

\bibitem{P71}
Pandharipande, V.~R. 1971, Nucl. Phys., A174, 641

\bibitem{PS75a}
Pandharipande, V.~R., \& Smith, R.~A. 1975a, Nucl.
Phys., A237, 507

\bibitem{PS75b}
---------. 1975b, Phys. Letters, 59B, 15

\bibitem{Stroh96c} Strohmayer, T., Zhang, W.,
Swank, J.\,H., Smale, A., Titarchuk, L., \& Day, C.
1996, ApJ, 469, L9

\bibitem{Tay92} Taylor, J.\,H. 1992, Phil. Trans. R.
Soc. Lond., 341, 117

\bibitem{Th93} Thorsett, S.\,E., Arzoumanian, Z.,
McKinnon, M.\,M., \& Taylor, J.\,H. 1993, ApJ, 405, L29

\bibitem{vdK97b} van der Klis, M. 1997b, these proceedings

\bibitem{W97}
Wijnands, R.\,A.\,D., van der Klis, M., van Paradijs,
J., Lewin, W.\,H.\,G., Lamb, F.\,K., Vaughan, B., \&
Kuulkers, E. 1997, ApJ, 479, L141

\bibitem{WFF88}
Wiringa, R.\,B., Fiks, V., \& Fabrocini, A. 1988,
Phys. Rev., C38, 1010

\bibitem{Z96}
Zhang, W., Lapidus, I., White, N.~E., \& Titarchuk,
L. 1996, ApJ, 469, L17

\bibitem{ZSS97}
Zhang, W., Strohmayer, T., \& Swank, J.~H. 1997,
ApJ, 482, L167

\end{thebibliography}
\end{document}